\documentclass[twocolumn]{aastex631}

\shorttitle{A complex dust morphology in Mrk~876}
\shortauthors{Landt et al.}
\accepted{2023 February 2 to \apj}

\begin{document}

\def\la{\mathrel{\hbox{\rlap{\hbox{\lower4pt\hbox{$\sim$}}}\hbox{$<$}}}}
\def\ga{\mathrel{\hbox{\rlap{\hbox{\lower4pt\hbox{$\sim$}}}\hbox{$>$}}}}

\font\sevenrm=cmr7
\def\OIII{[O~{\sevenrm III}]}
\def\FeII{Fe~{\sevenrm II}}
\def\FeIIf{[Fe~{\sevenrm II}]}
\def\SIII{[S~{\sevenrm III}]}
\def\HeI{He~{\sevenrm I}}
\def\HeII{He~{\sevenrm II}}
\def\NeV{[Ne~{\sevenrm V}]}
\def\OIV{[O~{\sevenrm IV}]}

\def\iraf{{\sevenrm IRAF}}
\def\mpfit{{\sevenrm MPFIT}}
\def\galfit{{\sevenrm GALFIT}}
\def\prepspec{{\sevenrm PrepSpec}}
\def\mapspec{{\sevenrm mapspec}}
\def\cream{{\sevenrm CREAM}}
\def\javelin{{\sevenrm JAVELIN}}
\def\cloudy{{\sevenrm CLOUDY}}
\def\banzai{{\sevenrm BANZAI}}
\def\orac{{\sevenrm ORAC}}
\def\demc{{\sevenrm DEMC}}

\def\gp{\mathcal{GP}}

\title{A complex dust morphology in the high-luminosity AGN Mrk~876}

\author[0000-0001-8391-6900]{Hermine Landt}
\affiliation{Centre for Extragalactic Astronomy, Department of Physics, Durham University, South Road, Durham, DH1 3LE, UK}

\author{Jake A. J. Mitchell}
\affiliation{Centre for Extragalactic Astronomy, Department of Physics, Durham University, South Road, Durham, DH1 3LE, UK}

\author[0000-0003-1810-0889]{Martin J. Ward}
\affiliation{Centre for Extragalactic Astronomy, Department of Physics, Durham University, South Road, Durham, DH1 3LE, UK}

\author{Paul Mercatoris}
\affiliation{Max Planck Institut f\"ur Astronomie, K\"onigstuhl 17, D-69117 Heidelberg, Germany}

\author[0000-0003-4291-2078]{J\"org-Uwe Pott}
\affiliation{Max Planck Institut f\"ur Astronomie, K\"onigstuhl 17, D-69117 Heidelberg, Germany}

\author[0000-0003-1728-0304]{Keith Horne}
\affiliation{SUPA Physics and Astronomy, University of St. Andrews, Fife, KY16 9SS, UK}

\author[0000-0002-6733-5556]{Juan V.\ Hern\'{a}ndez Santisteban}
\affiliation{SUPA Physics and Astronomy, University of St. Andrews, Fife, KY16 9SS, UK}

\author{Daksh Malhotra}
\affiliation{SUPA Physics and Astronomy, University of St. Andrews, Fife, KY16 9SS, UK}
\affiliation{Department of Physics, University of Alberta, 4-181 CCIS, Edmonton, AB T6G 2E1, Canada}

\author[0000-0002-8294-9281]{Edward M. Cackett}
\affiliation{Wayne State University, Department of Physics \& Astronomy, 666 W Hancock St, Detroit, MI 48201, USA}

\author[0000-0002-2908-7360]{Michael R. Goad}
\affiliation{Department of Physics and Astronomy, University of Leicester, University Road, Leicester, LE1 7RH, UK}

\author[0000-0003-0607-1136]{Encarni Romero Colmenero}
\affiliation{Southern African Large Telescope (SALT), P.O. Box 9, Observatory 7935, Cape Town, South Africa}
\affiliation{The South African Astronomical Observatory (SAAO), P.O. Box 9, Observatory 7935, Cape Town, South Africa}

\author[0000-0003-2662-0526]{Hartmut Winkler}
\affiliation{Department of Physics, University of Johannesburg, P.O. Box 524, 2006 Auckland Park, Johannesburg, South Africa}

\correspondingauthor{Hermine Landt}
\email{hermine.landt@durham.ac.uk}

\begin{abstract}

Recent models for the inner structure of active galactic nuclei (AGN) advocate the presence of a radiatively accelerated, dusty outflow launched from the outer regions of the accretion disk. Here we present the first near-infrared (near-IR) variable (rms) spectrum for the high-luminosity, nearby AGN Mrk~876. We find that it tracks the accretion disk spectrum out to longer wavelengths than the mean spectrum due to a reduced dust emission. The implied outer accretion disk radius is consistent with the infrared results predicted by a contemporaneous optical accretion disk reverberation mapping campaign and much larger than the self-gravity radius. The reduced flux variability of the hot dust could be either due to the presence of a secondary, constant dust component in the mean spectrum or introduced by the destructive superposition of the dust and accretion disk variability signals or some combination of both. Assuming thermal equilibrium for optically thin dust, we derive the luminosity-based dust radius for different grain properties using our measurement of the temperature. We find that in all cases considered the values are significantly larger than the dust response time measured by IR photometric monitoring campaigns, with the least discrepancy present relative to the result for a wavelength-independent dust emissivity law, i.e. a blackbody, which is appropriate for large grain sizes. This result can be well explained by assuming a flared, disk-like structure for the hot dust.

\end{abstract}


\section{Introduction}

Most active galactic nuclei (AGN) display prominent infrared (IR) emission in their spectral energy distributions (SEDs), which can be attributed to thermal dust radiation. This central dusty structure is commonly assumed to be optically thick and have a toroidal geometry aligned with the plane of the accretion disk. The latter requirement was mainly derived from the need of an equatorial obscurer in AGN in order to account for observations of broad emission lines in the polarized, scattered light of many type~2 AGN and the relative numbers of type~1 and type~2 AGN in complete samples \citep[see reviews by][]{Law87, Ant93, Netzer15, Lyu22a}. A warped, dusty disk could be a viable alternative to the dusty torus, as first proposed by \citet{Phi89b}. The extended dusty torus emits over a large range of IR wavelengths, with the hottest dust observed in the near-IR believed to be located closest to the central supermassive black hole, whereas warm dust (of a few 100~K) is observed in the mid-IR and expected to be located on average further out \citep{Weed05, Buch06, L10b, Lyu18, Brown19}. Often, cold dust (of a few 10~K) emitting at far-IR wavelengths is also observed in AGN, although this component is most likely heated by young stars in the host galaxy rather than by the ultraviolet (UV) and optical emission of the central accretion disk \citep{Kirk15}.

AGN are ideal laboratories to investigate the chemical composition and grain properties of astrophysical dust. Their UV/optical luminosities are usually high enough to heat the circumnuclear dust to sublimation temperatures. If these highest values can be observed, they can in principle constrain the chemistry since different species condense out of the gas phase in different environmental conditions. Dust temperatures were measured with simultaneous photometry at several near-IR wavelengths in a handful of sources \citep{Clavel89, Glass04, Schnuelle13, Schnuelle15}, but has only come of age with the availability of efficient near-IR cross-dispersed spectrographs. \citet{L11a} and \citet{L14} derived dust temperatures from such spectroscopy for the largest sample of type~1 AGN so far ($\sim 30$ sources). Their measurements yielded a very narrow temperature distribution with an average value of $T \sim 1400$~K. This result indicated that, if the hot dust is indeed at sublimation, it is composed of only silicate dust grains and so an oxygen-rich environment from which the dust formed. However, if carbon is present, then the dust is {\it not} heated to close to sublimation, since carbonaceous dust, e.g. graphite, can survive up to $T \sim 2000$~K \citep{Sal77}.

The range in temperatures within the dusty torus roughly translates to a range in radius, whereby mainly the mid-IR emitting dust can be spatially resolved with current instruments. High-angular-resolution mid-IR imaging \citep{Ramos11} and mid-IR interferometry \citep[e.g.][]{Tri09, Bur13} of a dozen nearby and bright AGN delivered useful upper limits on the extent of the dusty torus (of up to a few parsec). Such spatially resolved observations have also revealed a significant warm dust component perpendicular to the plane of the accretion disk, referred to as 'polar dust' \citep{Hoenig13, Tri14, Isbell22, Lyu22b, Gamez22}, which was also evident from SED studies \citep{L10b, Isbell21}. GRAVITY, the near-IR interferometric instrument at the Very Large Telescope (VLT), has started to resolve the innermost and hottest part of the central dusty structure in some nearby, luminous sources \citep{gravity20} and further progress is expected from an upgrade in sensitivity to GRAVITY$+$. But for most AGN knowledge about the inner dust radius is most efficiently obtained through dust reverberation, which is a technique that measures the time response of the dust to the variable, irradiating accretion disk flux. For $\sim 60$ AGN, the hot dust radius was determined by photometric campaigns, often coordinated at optical and near-IR wavelengths \citep[e.g.][]{Clavel89, Nel96, Okn01, Glass04, Sug06, Schnuelle13, Kosh14, Schnuelle15, Vazquez15, Min19, Lyu19}. In general, observed dust response times follow a luminosity-radius relationship with a slope similar to that for the broad emission line region (BLR), indicating a roughly constant hot dust flux and so narrow hot dust temperature distribution. However, dust radii measured via reverberation are often smaller (by factors of a few) than dust radii measured by interferometry or estimated from the SED assuming thermal equilibrium \citep{Kish07, Nen08a, L14}. A possible interpretation for this finding is that the dust has a bowl-shaped geometry caused by the anisotropy of the accretion disk emission irradiating it \citep{Kaw10, Kaw11}. In such a geometry, the hot dust located in the plane of the accretion disk is placed further in than the bulk of the dust and so is expected to dominate the reverberation signal.

Progress in technology and flexible scheduling have now made spectroscopic near-IR monitoring campaigns feasible. \citet{L19} presented the first such program. Using a near-IR cross-dispersed spectrograph with a relatively wide wavelength coverage that extends partially into the optical they showed that such a campaign allows in low-redshift AGN: (i) the monitoring of a large portion of the hot dust SED that readily gives the dust temperature and its evolution; (ii) the separation of the accretion disk emission from that of the hot dust, which is a considerable source of error in photometric campaigns; (iii) the simultaneous determination of luminosity-based and response-weighted dust radii; (iv) the construction of the variable (rms) near-IR spectrum; and (v) the study of the variability of emission lines formed in the BLR and the coronal line region. Their study of the hot dust in NGC~5548 found that a single component dominated both the mean emission and variations, with the dust reponse time and the luminosity-based dust radius being consistent with each other only if a blackbody emissivity was assumed. This result constrained the dust grain size to a few $\mu$m. The temperature and its variability indicated carbonaceous dust well below the sublimation threshold undergoing a heating and cooling process in response to the variable UV/optical accretion disk flux irradiating it. Most importantly, the dust reverberation signal showed tentative evidence for a second hot dust component most likely associated with the accretion disk. The existence of such dust is a prerequisite for the recent models of the AGN structure proposed by \citet{Czerny17} and \citet{Baskin18}, which explain both the BLR and dusty torus as part of the same outflow launched from the outer regions of the accretion disk by radiation pressure on dust, preferentially on carbonaceous dust since it has a higher opacity than silicate dust.

Here we present results from a near-IR spectroscopic monitoring campaign on Mrk~876 conducted between 2016 May and 2017 Jul. A contemporaneous optical photometric monitoring campaign in several bands was conducted between 2016 Mar and 2019 May and \citet{Miller22} have recently presented the accretion disk reverberation results from these data. The structure of our paper is as follows. In Sections~2, we discuss the science target and give the details of the near-IR observations and data reduction in Section~3. In Sections~4 and 5, we derive the luminosity-based dust radius and the variable (rms) spectrum, respectively. We discuss our main results in Section~6, where we seek a relation between the central dust structures observed in AGN and protoplanetary disks around young stars. Finally, in Section~7, we present a short summary and our conclusions.

\section{The science target} \label{target}

Mrk~876 (PG~1613$+$658) is one of the intrinsically most luminous AGN in the nearby Universe. At a redshift of $z=0.129$ it has an average $V$-band luminosity of $\sim 5 \times 10^{44}$~erg~s$^{-1}$ \citep{Bentz13}. This places it at the undersampled top end of the relationship between the hot dust radius and optical continuum luminosity presented by \citet{Kosh14} and \citet{Min19}. The latter study performed for Mrk~876 a co-ordinated optical and near-IR photometric campaign during the years \mbox{2003-2007} and measured a dust response time of $\tau = 320\pm34$, $327^{+42}_{-36}$ and $327^{+41}_{-35}$~light-days, assuming different values for the accretion disk power-law spectral slope when decomposing the flux in the $K$~band. A similar dust response time was obtained also by \citet{Lyu19}, who combined optical photometric light-curves from ground-based transient surveys with the {\it WISE} monitoring data for the years \mbox{2010-2018}. They estimated a dust response time of $\tau = 372\pm63$ and $390\pm37$~light-days for the $W1$ and $W2$ bands, respectively. Finally, \citet{Afa19} performed spectropolarimetric observations of Mrk~876 in 2015 and constrained the distance to the equatorial scattering region to $254\pm39$~light-days, which they assumed was the inner radius of the dusty torus.

Mrk~876 has a relatively high black hole mass, well-determined by optical reverberation campaigns of \mbox{$M_{\rm BH} = (2.2 \pm 1.0) \times 10^8~M_\odot$} \citep{Pet04, Bentz15}, transformed from the measured virial product using a scaling factor of $f=4.3$ \citep{Grier13b}. For this black hole mass, the corresponding gravitational radius is $r_{\rm g} = 3.2 \times 10^{13}$~cm = 0.013~light-days, where $r_{\rm g}=G M_{\rm BH}/c^2$, with $G$ the gravitational constant and $c$ the speed of light. The corresponding Eddington luminosity is $L_{\rm edd} = 2.7 \times 10^{46}$~erg~s$^{-1}$. The optical emission-line spectrum of Mrk~876 is of the inflected type, i.e. the broad line profiles have clearly discernible broad- and narrow-line components. The hydrogen BLR radius has been recently measured by an optical spectroscopic reverberation mapping campaign during the years \mbox{2016-2021} to be in the range of $\sim 40 - 50$~light-days, based on the response time of the optical Balmer line H$\beta$ \citep{Bao22}.

Mrk~876 (J2000 sky coordinates \mbox{R.A. $16^h 13^m 57.2^s$}, \mbox{Decl. $+65^\circ 43' 10''$}) is observable only from the northern hemisphere. Its low redshift and high intrinsic luminosity make it sufficiently bright in the near-IR \citep[2MASS $J=13.2$~mag, $K_s=11.4$~mag;][]{2MASS} so that a high-quality spectrum can be achieved in a relatively short exposure time. We adopt the cosmological parameters \mbox{$H_0 = 70$~km~s$^{-1}$~Mpc$^{-1}$}, $\Omega_{\rm M}=0.3$, and $\Omega_{\Lambda}=0.7$, which give a luminosity distance to Mrk~876 of 605.2 Mpc and an angular scale at the source of 2302~pc per arcsec.

\section{The observations}

\subsection{The near-IR spectroscopy} \label{spectroscopy}

\begin{deluxetable*}{lccccrrrclc}
\tablecolumns{11}
\decimalcolnumbers
\tablecaption{\label{obslog} 
Gemini journal of observations}
\tablehead{
\colhead{Observation} & \colhead{exposure} & \colhead{airmass} & \colhead{aperture} & \colhead{PA} & \multicolumn{3}{c}{continuum S/N} & telluric star & \twocolhead{correction factor} \\
\colhead{Date} & \colhead{(s)} & & \colhead{(arcsec$^2$)} & \colhead{($^{\circ}$)} & \colhead{J} & \colhead{H} & \colhead{K} & \colhead{airmass} & \colhead{\prepspec} & \colhead{\mapspec}
}
\startdata
2016 May 25 & 8$\times$120 & 1.553 & 0.675$\times$1.25 & 143 &  77 & 112 & 180 & 1.591 & 1.01$\pm$0.03          & 1.12 \\
2016 Jun 16 & 8$\times$120 & 1.438 & 0.675$\times$1.11 & 165 & 104 & 217 & 105 & 1.392 & 0.71$\pm$0.04          & 0.83 \\
2016 Jul 15 & 8$\times$120 & 1.490 & 0.675$\times$0.95 & 146 & 124 & 154 & 201 & 1.401 & 1.11$^{+0.14}_{-0.12}$ & 1.12 \\
2016 Aug 4  & 8$\times$120 & 1.455 & 0.675$\times$1.09 & 158 &  87 & 115 & 212 & 1.422 & 0.99$^{+0.10}_{-0.09}$ & 1.18 \\
2017 Feb 24 & 8$\times$120 & 1.485 & 0.675$\times$1.28 & 190 & 105 & 245 & 145 & 1.372 & 1.23$\pm$0.05          & 1.31 \\
2017 Apr 4  & 8$\times$120 & 1.755 & 0.675$\times$1.27 & 174 & 105 & 139 & 187 & 1.706 & 0.97$\pm$0.05          & 1.08 \\
2017 Apr 16 & 8$\times$120 & 1.481 & 0.675$\times$1.08 & 201 & 118 & 162 & 160 & 1.416 & 1.15$\pm$0.03          & 1.21 \\
2017 May 4  & 6$\times$120 & 1.439 & 0.675$\times$1.26 & 187 & 109 &  93 & 165 & 1.362 & 0.88$^{+0.05}_{-0.04}$ & 0.87 \\
2017 Jun 3  & 8$\times$120 & 1.443 & 0.675$\times$1.18 & 174 &  96 & 102 & 104 & 1.414 & 0.99$^{+0.08}_{-0.07}$ & 1.07 \\
2017 Jul 5  & 8$\times$120 & 1.489 & 0.675$\times$1.40 & 174 &  72 & 110 & 194 & 1.423 & 1.43$\pm$0.04          & 1.63
\enddata

\tablecomments{The columns are: (1) Universal Time (UT) date of observation; (2) exposure time; (3) mean airmass; (4) extraction aperture; (5) slit position angle, where PA$=0^{\circ}$ corresponds to north-south orientation and is defined north through east; S/N in the continuum over $\sim 100$~\AA~measured at the central wavelengths of the (6) J, (7) H, and (8) K bands; (9) mean airmass for the telluric standard star; and multiplicative photometric correction factor determined using the narrow emission line~\SIII~$\lambda~9531$ within the (10) \prepspec~and (11) \mapspec~routines.}

\end{deluxetable*}

\begin{figure*} 
\centerline{
\includegraphics[scale=0.7]{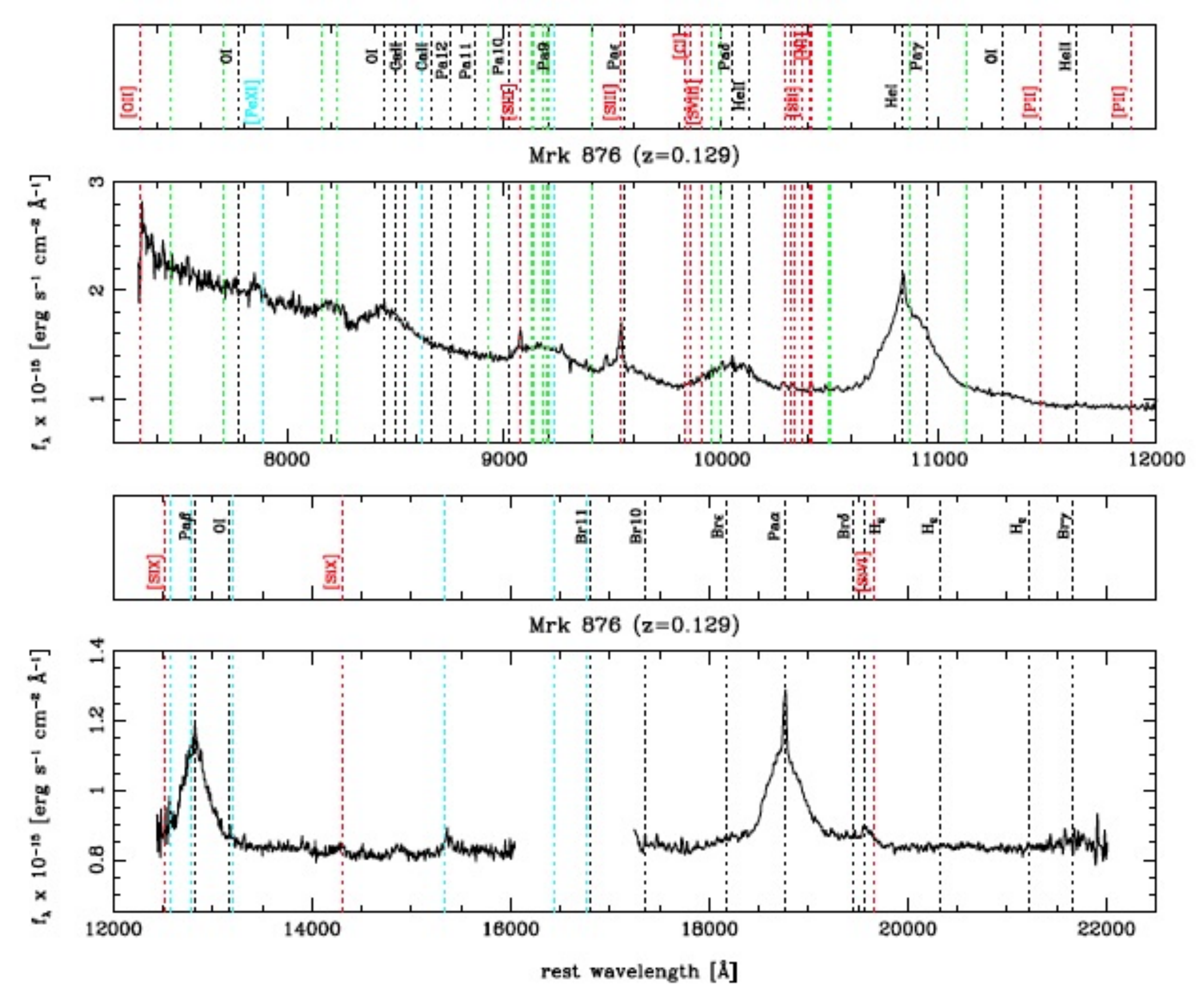}
}
\caption{\label{gnirsspec} Gemini GNIRS near-IR spectrum from 2016 May 25, shown as observed flux versus rest-frame wavelength. Emission lines listed in Table 4 of \citet{L08a} are marked by dotted lines and labeled; black: permitted transitions, green: permitted \FeII~multiplets (not labeled), red: forbidden transitions and cyan: forbidden transitions of iron (those of \FeIIf~not labeled).}
\end{figure*}

We observed Mrk~876 between 2016 May and 2017 Jul at the Gemini North 8~m observatory on Maunakea, Hawaii, in queue mode (Program ID: \mbox{GN-2016A-FT-27}, \mbox{GN-2017A-Q-41}) using the Gemini Near-Infrared Spectrograph \citep[GNIRS;][]{gnirs}. We obtained roughly one spectrum per month, except for the gap period of about five months (Sep to Jan) when Mrk~876 is not observable, resulting in a total of 10 near-IR spectra. Table \ref{obslog} lists the journal of observations. The sky was clear and the seeing was very good ($\sim 0.4\arcsec - 0.6\arcsec$) for all nights except 2016 Jun 16, 2017 Feb 24 and Jul 5, when some clouds were present and the seeing was variable. We used the cross-dispersed mode with the short camera at the lowest spectral resolution (31.7~l~mm$^{-1}$ grating), thus covering the entire wavelength range of $0.85-2.5$~$\mu$m without inter-order contamination. We chose a slit of $0.675\arcsec \times 7\arcsec$, which we oriented at the parallactic angle. This set-up gives an average spectral resolution of full width at half-maximum (FWHM) $\sim 400$~km~s$^{-1}$. The on-source exposure time of $8 \times 120$~s ensured that we obtained spectra with a high signal-to-noise ratio ($S/N$) in order to reliably measure the emission-line profiles. Since the source is not extended in the near-IR, we nodded along the slit in the usual ABBA pattern. The observations were taken on average at an airmass of $\sec z \sim 1.5$, which is close to the minimum value achievable for this source from Hawaii.

After or before the science target, we observed the nearby (in position and airmass) F0~V star HIP~79087 (HD~145710) that has accurate near-IR magnitudes. We used this standard star to correct our science spectrum for telluric absorption and for flux calibration. For the flux calibration we assumed that its continuum emission can be approximated by a blackbody with an effective temperature of $T_{\rm eff}=6769$~K \citep{pastel}. Flats and arcs were taken after the science target.

We reduced the data using the Gemini/IRAF package with GNIRS specific tools \citep{gnirssoft}. The data reduction steps included preparation of calibration and science frames, processing and extraction of spectra from science frames, wavelength calibration of spectra, telluric correction and flux-calibration of spectra, and merging of the different orders into a single, continuous spectrum. The spectral extraction width was adjusted interactively for the telluric standard star and the science source to include all the flux in the spectral trace. A local averaged background flux was fitted and subtracted from the total source flux. The final spectrum was corrected for Galactic extinction using the IRAF task \mbox{\sl onedspec.deredden} with an input value of $A_{\rm V}=0.005$, which we derived from the Galactic hydrogen column densities published by \citet{DL90}. In Fig. \ref{gnirsspec}, we show the spectrum from 2016 May 25 as a representative example.

\subsection{The complementary photometry} \label{photometry}

Mrk~876 was monitored with the 1~m robotic telescope network of the Las Cumbres Observatory \citep[LCOGT;][]{LCOGT} as part of the 2014 AGN Key Project almost daily in four bands ($g$, $r$, $i$ and $z_s$) between 2016 Feb 12 and 2020 Oct 10, but with the $z_s$ band switched off in the period of 2017 Apr 4 to 2018 Jun 17. We made use here of the light-curves in the $g$ and $z_s$ filters, which have a wavelength width of 1500~\AA~and 1040~\AA~around their central wavelength of 4770~\AA~and 8700~\AA, respectively. The LCOGT Sinistro cameras have a FOV of $26.5\arcmin \times 26.5\arcmin$ and a plate scale of $0.389\arcsec$ per pixel.

The frames were first processed by LCOGT's \mbox{\banzai}~pipeline \citep{banzai} in the usual way (bias and dark subtraction, flat-fielding correction and cosmic ray rejection) and were subsequently analysed with the custom-made pipeline described in detail by \citet{Hernandez20}. In short, after performing aperture photometry with a diameter of $7\arcsec$ and subtracting a background model, stable light-curves were produced by constructing a curve of growth using the standard stars on each individual frame and measuring the correction factors required to bring all different light-curves to a common flux level. A colour-correction and correction for atmospheric extinction were applied before the photometric calibration. Finally, an image zero-point calibration was performed at each epoch based on comparison stars in the field. In Fig. \ref{lcurves} (top two panels), we display the $g$ and $z_s$ light curves, with the latter overlapping in wavelength with the GNIRS near-IR spectrum.

\subsection{The absolute spectral flux scale}

\begin{figure}
\centerline{
\includegraphics[scale=0.3]{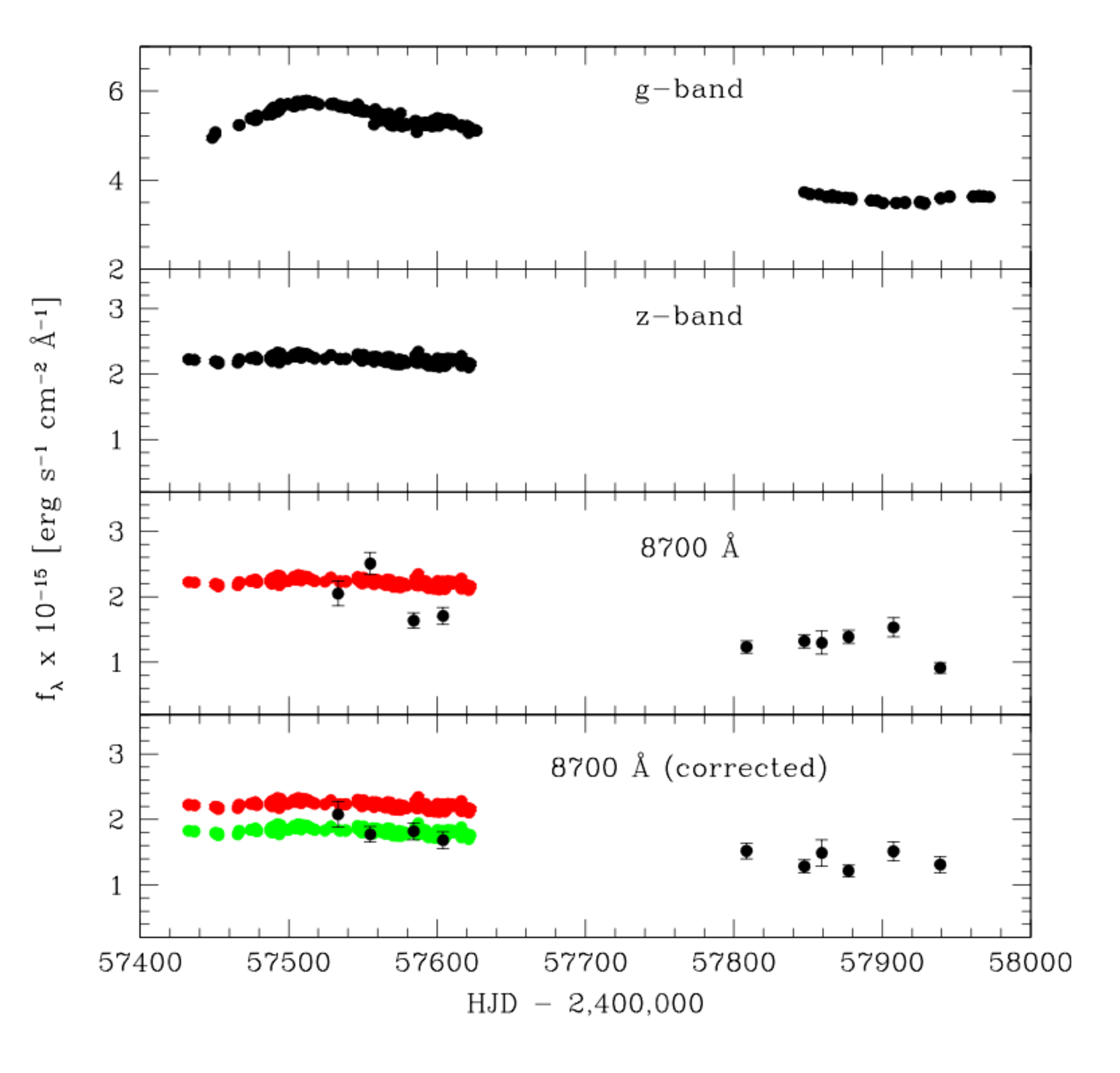}
}
\caption{\label{lcurves} Top two panels: LCOGT $g$ and $z_s$~band light-curves during the period 2016/17. Bottom two panels: \mbox{Gemini} GNIRS near-IR spectral light-curve around the observed wavelength of $8700$~\AA~(black filled circles), both original and corrected using \prepspec~photometric correction factors based on the \SIII~$\lambda 9531$~line, versus the $z_s$ band light-curve (red circles). The $z_s$ band light-curve corrected for a constant host galaxy contribution to match the near-IR spectral fluxes is also shown (green filled circles).}
\end{figure}

In order to derive light-curves, construct meaningful mean and variable (rms) spectra and estimate luminosities, we must achieve an accurate absolute flux calibration of the spectra. Similar to the approach chosen by optical spectroscopic reverberation campaigns to achieve this, we base our absolute spectral flux scale on a strong narrow emission line from a forbidden transition. Such emission lines are expected to remain constant during the campaign, since they are produced in gas that is located at distances of pc-to-kpc scales from the ionising source and has a number density low enough for recombination timescales to be large. We have chosen the \SIII~$\lambda 9531$~line since it is the strongest narrow forbidden emission line in the near-IR. It is blended with the Pa$\epsilon$~broad emission line, but can be easily separated from it since this hydrogen line is usually weak. Imaging studies of the \SIII~$\lambda 9531$~line have been performed in a few nearby AGN using near-IR integral-field spectrographs and the gas forming this line was found to be extended on moderate scales of up to a few 100~pc \citep[e.g.][]{Storchi09, Fischer15, Fischer17}. Our slit size of $0.675\arcsec$ centered on the nucleus corresponds to a radial extent of $\sim 800$~pc, which is sufficiently large to ensure that all the \SIII~line flux is enclosed in the spectral extraction aperture.

We have used two routines to determine the photometric correction factors for our observations, the \prepspec~routine developed by Keith Horne \citep[last described in][]{Horne20} and the \mapspec~routine of \citet{Faus17}. In short, \prepspec~models both the emission lines and the total continuum and subsequently matches the profiles of selected narrow emission lines in the spectra in order to derive the time-dependent scaling factor. It assumes that the majority of the spectra are photometric and so holds the median photometric correction factor at one. The \mapspec~routine constructs a reference spectrum by averaging the highest quality spectra, e.g., those spectra that have the highest S/N and/or were observed under the best weather conditions. Then, the profile of the chosen narrow emission line is matched between the reference spectrum and the individual spectra. The spectral scaling factors resulting from the two routines are listed in Table \ref{obslog}. With the exception of the last observational epoch, the results from \prepspec~and \mapspec~are consistent within the errors, but there is a trend for the former correction factors to be lower than the latter. In the following, we use the \prepspec~correction factors, which span a range of $\sim 1-40\%$, with an average value of $\sim 10\%$. In Fig.~\ref{lcurves} (bottom two panels), we show the GNIRS near-IR spectral continuum flux in the observed wavelength range of the LCOGT $z_s$ band filter centred at $8700$~\AA, both original and corrected. If we subtract from the $z_s$~band light-curve a constant host galaxy flux contribution of $\sim 20\%$ of the total flux, it matches very well all four of the corrected near-IR spectral flux measurements in the time of overlap. For the later epochs, the $g$~band light-curve shows a downward trend in flux, which is also reproduced.

\section{The luminosity-based dust radius} \label{lumradius}

\begin{figure}
\centerline{
\includegraphics[angle=-90, scale=0.32]{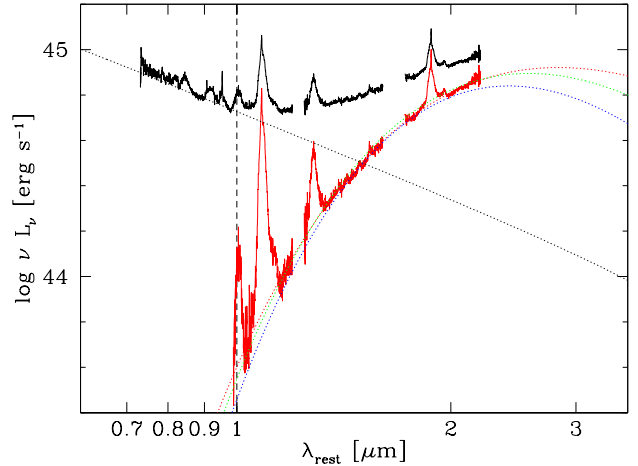}
}
\caption{\label{irsed} Gemini GNIRS near-IR spectrum from 2016 May 25 shown as luminosity versus rest-frame wavelength (black spectrum). We decomposed the continuum into two components, namely, an accretion disk spectrum that approximates the wavelength range $\la 1~\mu$m (black dotted curve) and still dominates at $1~\mu$m (vertical dashed line) and hot dust emission (red spectrum). We fitted the hot dust continuum with a blackbody spectrum (red dotted curve) and modified blackbody spectra for carbon and silicate dust (green and blue dotted curves, respectively) with resulting best-fit temperatures as listed in Table \ref{lumradiustab}.}
\end{figure}

\begin{deluxetable*}{lclcclcclcc}
\tablecolumns{11}
\decimalcolnumbers
\tablecaption{\label{lumradiustab} 
Physical parameters for the calculation of luminosity-based dust radii}
\tablehead{
\colhead{Observation} & \colhead{accretion} & \multicolumn{3}{c}{blackbody} & \multicolumn{3}{c}{silicate dust} & \multicolumn{3}{c}{carbon dust} \\
\colhead{Date} & \colhead{disk} & \multicolumn{3}{c}{($\beta=0$, $\langle Q^{\rm em} \rangle=1$)} 
& \multicolumn{3}{c}{($\beta=-1$, $\langle Q^{\rm em} \rangle=0.0210$)} & 
\multicolumn{3}{c}{($\beta=-2$, $\langle Q^{\rm em} \rangle=0.0875$)} \\
& \colhead{log $L_{\rm uv}$} & \colhead{$T_{\rm d}$} & \colhead{log $L_{\rm d}$} & \colhead{$R_{\rm d,lum}$} & \colhead{$T_{\rm d}$} & \colhead{$L_{\rm d}$} & \colhead{$R_{\rm d,lum}$} & \colhead{$T$} & \colhead{$L_{\rm d}$} & \colhead{$R_{\rm d,lum}$} \\
& \colhead{(erg/s)} & \colhead{(K)} & \colhead{(erg/s)} & \colhead{(lt-days}) & \colhead{(K)} & \colhead{(erg/s)} & \colhead{(lt-days)} & \colhead{(K)} & \colhead{(erg/s)} & \colhead{(lt-days)}}
\startdata
2016 May 25 & 46.22 & 1297$\pm$13 & 45.55 & 554 & 1126$\pm$11 & 44.73 & 5070 &  996$\pm$8 & 45.28 & 3175 \\
2016 Jun 16 & 46.19 & 1305$\pm$9  & 45.71 & 681 & 1133$\pm$7  & 44.84 & 6232 & 1001$\pm$5 & 45.70 & 3912 \\
2016 Jul 15 & 46.18 & 1325$\pm$11 & 45.51 & 468 & 1148$\pm$9  & 44.75 & 4298 & 1013$\pm$7 & 45.59 & 2704 \\
2016 Aug 4  & 46.14 & 1311$\pm$9  & 45.54 & 494 & 1136$\pm$7  & 44.73 & 4543 & 1002$\pm$6 & 45.67 & 2861 \\
2017 Feb 24 & 46.12 & 1324$\pm$10 & 45.53 & 408 & 1149$\pm$7  & 44.76 & 3737 & 1015$\pm$6 & 45.17 & 2346 \\
2017 Apr 4  & 46.01 & 1334$\pm$10 & 45.55 & 421 & 1158$\pm$7  & 44.78 & 3852 & 1023$\pm$5 & 45.63 & 2418 \\
2017 Apr 16 & 46.04 & 1303$\pm$8  & 45.52 & 402 & 1136$\pm$6  & 44.72 & 3651 & 1007$\pm$5 & 45.19 & 2276 \\
2017 May 4  & 46.00 & 1308$\pm$14 & 45.61 & 458 & 1135$\pm$11 & 44.43 & 4199 & 1003$\pm$8 & 45.57 & 2634 \\
2017 Jun 3  & 46.01 & 1267$\pm$13 & 45.55 & 466 & 1102$\pm$10 & 44.34 & 4254 &  975$\pm$8 & 45.53 & 2662 \\
2017 Jul 5  & 46.02 & 1288$\pm$12 & 45.43 & 342 & 1120$\pm$9  & 44.68 & 3124 &  991$\pm$7 & 45.03 & 1955
\enddata

\tablecomments{The columns are: (1) Universal Time (UT) date of observation; (2) total accretion disk luminosity; for a blackbody emissivity (3) dust temperature; (4) total dust luminosity and (5) dust radius; for an emissivity law appropriate for silicate dust with small grain sizes of $a \la 0.1~\mu$m (6) dust temperature; (7) total dust luminosity and (8) dust radius; for an emissivity law appropriate for carbon dust with small grain sizes of $a \la 0.1~\mu$m (9) dust temperature; (10) total dust luminosity and (11) dust radius.}

\end{deluxetable*}

In order to be able to derive luminosity-based dust radii, we need to measure the dust temperature and estimate the UV/optical accretion disk luminosity that heats the dust. The relatively large wavelength range of the cross-dispersed near-IR spectra covers in Mrk~876 simultaneously about half the hot dust SED and a considerable part of the accretion disk spectrum, which is expected to dominate the total continuum flux up to $\sim 1~\mu$m \citep{L11b, L11a}. Therefore, we decomposed the spectral continuum into two components following the approach described in \citet{L19} and exemplified in Fig.~\ref{irsed}. We note that, since we used a relatively small spectral aperture, the contribution of host galaxy light to the total observed continuum flux is negligible in this luminous AGN. In short, we have first approximated the rest-frame wavelength range of $\la 1~\mu$m with the spectrum of a standard accretion disk, which we have subsequently subtracted from the total spectrum. For the calculation of the accretion disk spectrum we adopted the black hole mass given in Section \ref{target} and the accretion rate was obtained directly from the a scaling to the data. Furthermore, we assumed that the disk is relatively large and extends out to $r_{\rm out}=10^4 r_{\rm g}$. We then fitted the resultant hot dust spectrum at wavelengths $>1~\mu$m with a blackbody, representing emission by large dust grains, and also with two blackbodies modified by a power-law of the form $Q_{\lambda} (a) \propto \lambda^{\beta}$, approximating with $\beta=-1$ and $\beta=-2$ the emissivity of sub-micron silicate and carbon dust grains, respectively \citep[][see their Fig. 8]{L19}. We note in Fig.~\ref{irsed} the excellent agreement between the data and the standard accretion disk spectrum that we assumed here, a result generally found for AGN with negligible contribution from host galaxy light \citep[e.g.][]{Korat99, L11a}. Table \ref{lumradiustab} lists the physical parameters derived from the spectral decomposition. The resultant average temperatures and the error on the mean are \mbox{$\langle T \rangle = 1306\pm6$}, 1134$\pm$5 and 1033$\pm$4~K for an emissivity law with $\beta=0$, $-1$ and $-2$, respectively.

As in \citet{L19}, we then calculated luminosity-based dust radii, $R_{\rm d,lum}$, from the best-fit dust temperatures assuming radiative equilibrium between the luminosity of the irradiating source and the dust:

\begin{equation}
\label{Stefan-Boltz}
\frac{L_{\rm uv}}{4 \pi R_{\rm d,lum}^2} = 4 \sigma T^4 \langle Q^{\rm em} \rangle,
\end{equation}

\noindent
where $\sigma$ is the Stefan-Boltzmann constant and $\langle Q^{\rm em} \rangle$ is the Planck-averaged value of $Q_{\lambda} (a)$. We have approximated $L_{\rm uv}$ with the accretion disc luminosity at each epoch, as given in Table \ref{lumradiustab}, and have used for the Planck-averaged emission efficiencies in the case of \mbox{$\beta=-1$} a value of \mbox{$\langle Q^{\rm em} \rangle=0.0210$} appropriate for silicates of \mbox{$T=1259$~K} and \mbox{$a=0.1~\mu$m} \citep{Laor93} and in the case of \mbox{$\beta=-2$} a value of \mbox{$\langle Q^{\rm em} \rangle=0.0875$} appropriate for graphite of \mbox{$T=1000$~K} and \mbox{$a=0.1~\mu$m} \citep{Draine16}. The average luminosity-based rest-frame dust radii and the error on the mean are \mbox{$\langle R_{\rm d,lum} \rangle = 469\pm30$}, 4296$\pm$273 and 2694$\pm$172~light-days in the case of a blackbody, and small-grain silicate and carbon dust, respectively.

\section{The variable near-IR spectrum} \label{rms}

\begin{figure*}
\centerline{
\includegraphics[angle=-90, scale=0.55]{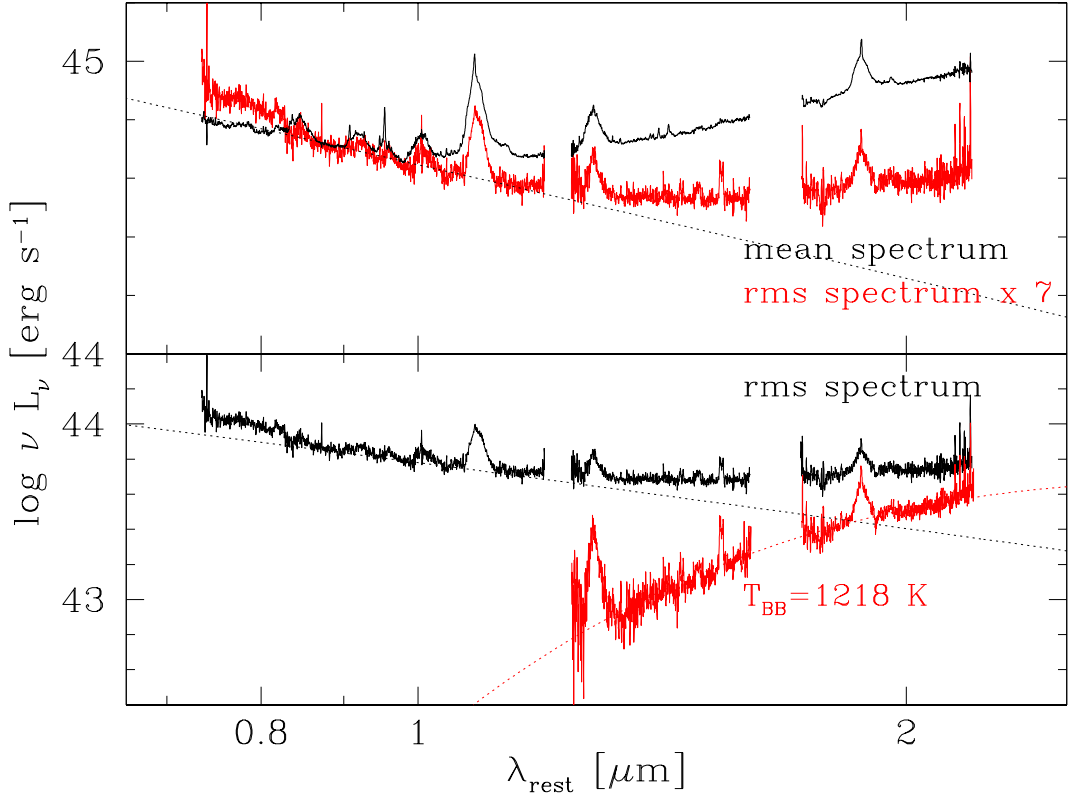}
}
\caption{\label{rmsspec} Top panel: The mean (black) and variable near-IR spectrum (red) for our campaign normalised at rest-frame $0.9~\mu$m. The spectrum of a standard accretion disk (black dotted line) is traced to $\sim 1.2~\mu$m in the variable component and only to $\sim 1~\mu$m in the mean spectrum. Bottom panel: The spectral decomposition of the variable near-IR spectrum (black) into an accretion disk spectrum (black dotted line) and a hot dust component (red spectrum) yields a blackbody temperature of $T \sim 1200$~K for the latter (red dotted curve).}
\end{figure*}

\begin{figure}
\centerline{
\includegraphics[angle=-90, scale=0.32]{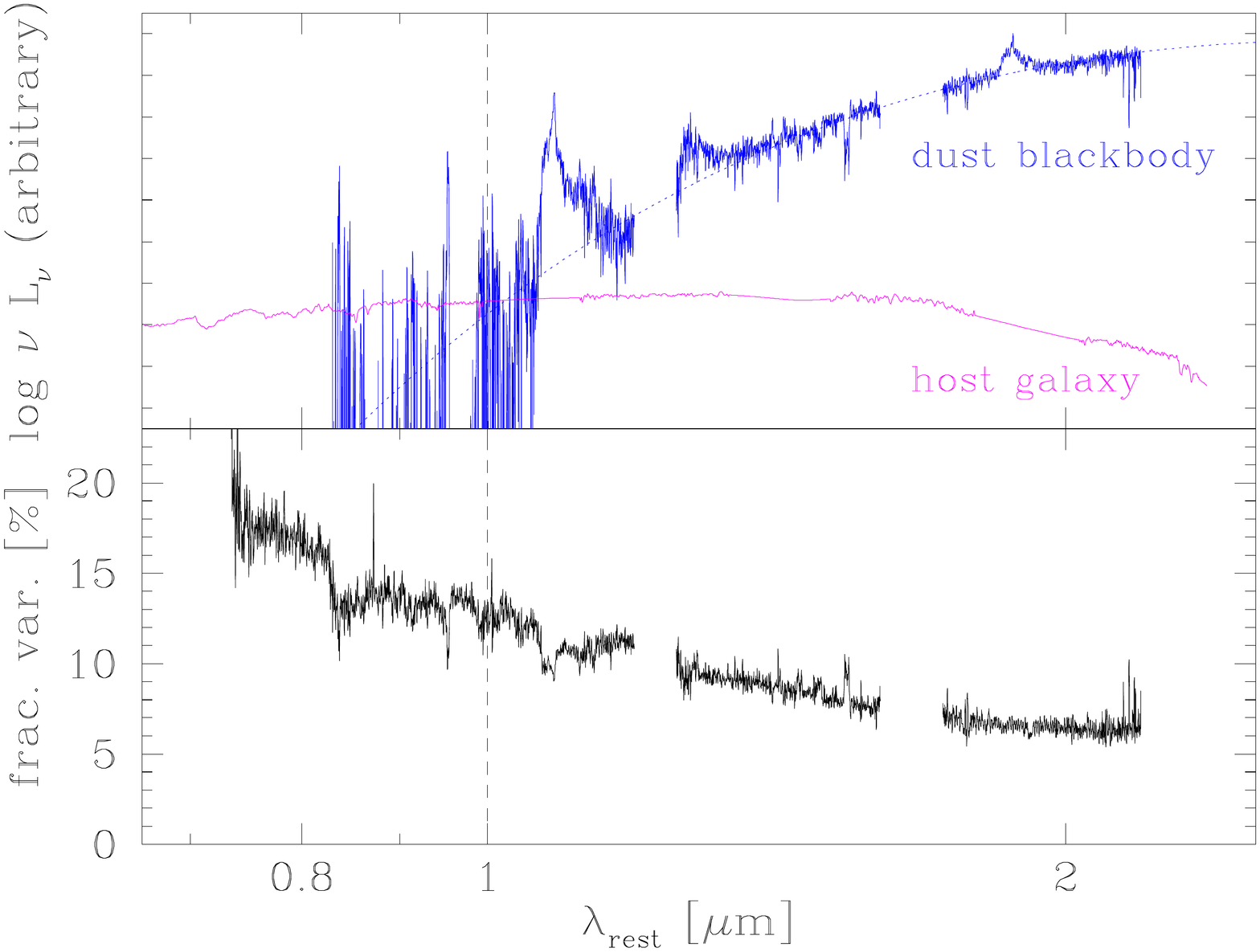}
}
\caption{\label{fracvar} Top panel: The flux difference between the normalised mean and variable spectra (blue) can be attributed to an additional (non-variable) dust component with a blackbody temperature of $T \sim 1300$~K (blue dotted line). The template of an elliptical host galaxy is also shown (magenta). Bottom panel: The fractional amplitude of the variations as a percentage of the mean flux versus the rest-frame wavelength.}
\end{figure}

We calculated the mean and variable (rms) near-IR spectra following \citet{Pet04}. Fig.~\ref{rmsspec} (top panel) shows them in comparison. In order to be able to compare the chromatic shape of the two spectra, we rescaled the rms spectrum by a factor of $\sim 7$ in order to match the mean optical flux at $0.9~\mu$m. Assuming a simple flux variability, scaling with the total flux would lead to comparable shapes, but we observe this similarity only in the spectral part dominated by the accretion disk. Instead two spectral continuum features become apparent at wavelengths beyond $\sim 1~\mu$m. First, the rms spectrum tracks the accretion disk spectrum out to longer wavelengths ($\sim 1.2~\mu$m) than the mean spectrum ($\sim 1~\mu$m). This is an important finding, since we do not know the full extent of accretion disks, mainly because their infrared spectrum is overwhelmed by the emission from the hot dust. Infrared polarimetry can reveal the disk near-IR continuum, but only for the brightest AGN \citep{Kish08}. In order for the accretion disk spectrum to dominate the emission at $\sim 1~\mu$m, as observed in the mean spectrum and established by the near-IR radius-luminosity relationship \citep{L11b}, the inferred disk outer radius is $r_{\rm out} \sim 1500\,r_{\rm g} \sim 20$~light-days. Then, observing the optical variable component in the rms spectrum out to $\sim 1.2~\mu$m and assuming it is the accretion disk implies a value of $r_{\rm out} \sim 1900\,r_{\rm g} \sim 25$~light-days. This radius is much larger than the self-gravity radius, i.e. the radius where accretion disks are generally assumed to become unstable and fragment, which is $\sim 12$~light-days \citep{Lobban22}.

Secondly, the near-IR flux variability relative to the mean continuum flux at $\geq 1~\mu$m appears significantly lower than in the visible domain, which is dominated by the accretion disk only. Since the near-IR shows clearly the additional hot dust flux, this behaviour could be explained by a systematically reduced flux variability of the hot dust with respect to the accretion disk. This would explain why the accretion disk spectral slope is revealed out to longer wavelengths, a situation similar to that seen in the {\it mean} flux spectrum of the so-called ``hot-dust-poor'' quasars reported by \citet{Hao10}. A spectral decomposition of the rms and mean spectra, as performed in Section \ref{lumradius}, gives a flux ratio reduced by a factor of $\sim 3$ in the former relative to the latter. The ratio of the rms to the mean spectrum measures the fractional amplitude of the variations, i.e. the variable flux as a percentage of the mean flux, in dependence of wavelength. If the shape of the variable spectrum is identical to that of the mean spectrum, it means that the fractional variability is independent of wavelength. A reduction of the fractional variability at a given wavelength can be introduced by the presence of a constant component, which would increase the flux in the mean spectrum. Our observed variable near-IR spectrum for Mrk~876 shows a reduced fractional variability for the blackbody component only (Fig.~\ref{fracvar}, bottom panel). We visualize this difference spectrum for wavelengths $\ga 1~\mu$m, which can be well-fitted by a blackbody of temperature $T \sim 1300$~K, in Fig.~\ref{fracvar} (top panel). Since the host galaxy of Mrk~876 is unusually luminous \citep[e.g.][]{Bentz09}, we have also considered the case that some of the spectral difference between the mean and rms spectrum is induced by this constant component. In Fig.~\ref{fracvar} (top panel), we have included the template of an elliptical galaxy. It is clear that a significant contribution from this component is unlikely given that the flux difference between the mean and rms spectrum increases with wavelength, whereas the host galaxy spectral flux shows the opposite behaviour.

Therefore, the hot dust seems to be composed of at least two components with very different variability timescales or behaviours. The secondary dust component present in the mean spectrum appears constant in our campaign either because it is truly non-variable or because it varies on timescales not probed by our monitoring campaign. If such a constant dust component is indeed present, then the variable dust component does not seem to dominate the total dust emission and constitutes only $\sim 30\%$ of it. A spectral decomposition of the variable near-IR spectrum done as in Section \ref{lumradius} yields that its blackbody temperature is $T \sim 1200$~K (Fig.~\ref{rmsspec}, bottom panel) and so slightly lower than the average value of the total dust emission of $\langle T \rangle \sim 1300$~K.

As an alternative explanation of the observed rms near-IR spectrum in Fig.~\ref{rmsspec}, it is conceivable that the deficit in fractional variability at wavelengths \mbox{$\ga 1.2~\mu$m} is not due to the presence of a secondary, non-variable dust component, but rather introduced by anti-correlation of the variability of the near-IR accretion disk flux and the hot dust flux, with the latter varying in delayed response to the former. In such a case, we expect the total variability signal in the near-IR to be a superposition of the two variability signals and so the spectral shape of the rms spectrum to depend on the interplay between them. This superposition can then either enhance or weaken the total variance relative to the mean. In the following, we have simulated this situation with the \demc~formalism assuming that the secondary, variable near-IR component is the accretion disk itself. 

The \demc~algorithm was developed by \citet{Schnuelle15} to suit multi-band optical/near-IR photometric dust reverberation campaigns and it can effectively construct the rms spectrum over this entire wavelength range. The analysis is carried out in two steps. First, the structure function parameters for the accretion disk variability are derived from modelling the optical light-curve by maximum likelihood based on the method of interpolation and reconstruction of noisy, irregularly sampled data by \citet{Ryb92}. The covariance function of the underlying Gaussian Process (GP) model is assumed to be a power-law \citep[e.g.][]{Press92, SchmidtK10, Hern15} and the corresponding structure function can be written as:

\begin{equation}
V(|t-t'|) = A^2 \left( \frac{t-t'}{1 {\rm yr}} \right)^\gamma,
\end{equation}

\noindent
where $A$ is the amplitude of the variability on a one-year timescale and $\gamma$ is the gradient of this variability. Secondly, a model consisting of the sum of a power-law, which represents the accretion disk emission, and a single blackbody, which represents the hot dust emission, is fit to the simultaneous multi-band optical/near-IR data. This model is:

\begin{equation}
f_\lambda(\lambda,t,\mathbf{x}) = C_1 f_g(t) \lambda^{-\alpha} + C_2 \frac{2hc^2}{\lambda^5} \frac{1}{e^{(hc/\lambda k_B T(t))} -1}, 
\end{equation}

\noindent
where $f_g(t)$ and $\alpha$ are the variable accretion disk flux in the optical and the spectral power-law index, respectively, $h$ is the Planck constant and $k_B$ is the Boltzmann constant. The dust model temperature $T(t)$ evolves with time from an initial value $T_0$ as \mbox{$dT(t)/T(t) = \nu \cdot 0.25~df_g(t-\tau)/f_g(t-\tau)$} in response to the variable accretion disk flux $f_g(t-\tau)$. The accretion disk flux is time-shifted by $\tau$ and processed with an efficiency $\nu$ (with $0<\nu<1$). The posterior distribution of the vector consisting of six parameters $\mathbf{x} = (C_1, C_2, \nu, \alpha, T_0, \tau)$ is sampled by the Differential Evolution Markov Chain \citep[DEMC;][]{TerBraak06} algorithm, which is basically a MCMC algorithm with multiple chains run in parallel. The algorithm evaluates the posterior probability density function for each iteration step and for each chain given uniform priors within sensible limits.

\begin{figure*}
\centerline{
  \includegraphics[scale=0.4]{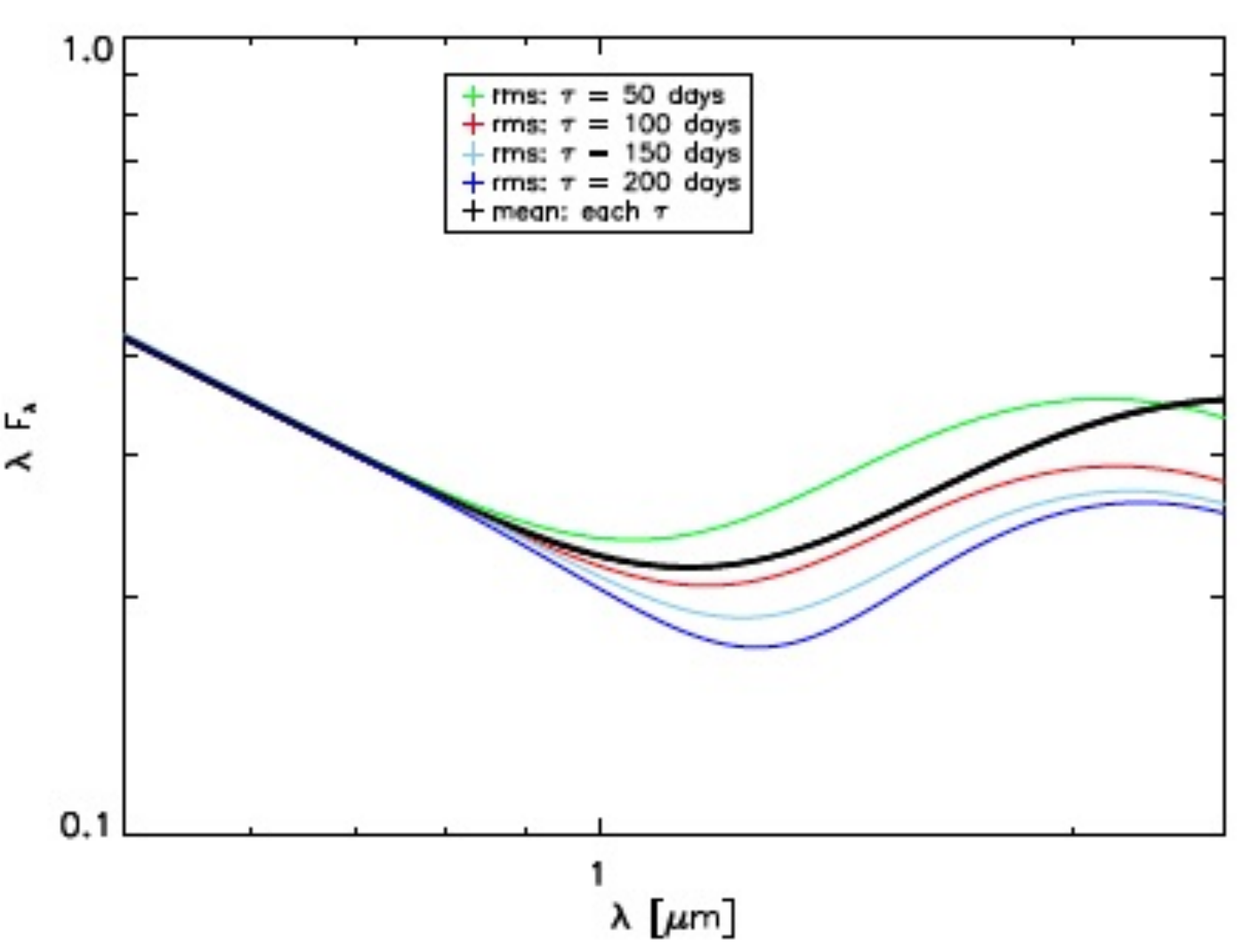}
  \includegraphics[scale=0.4]{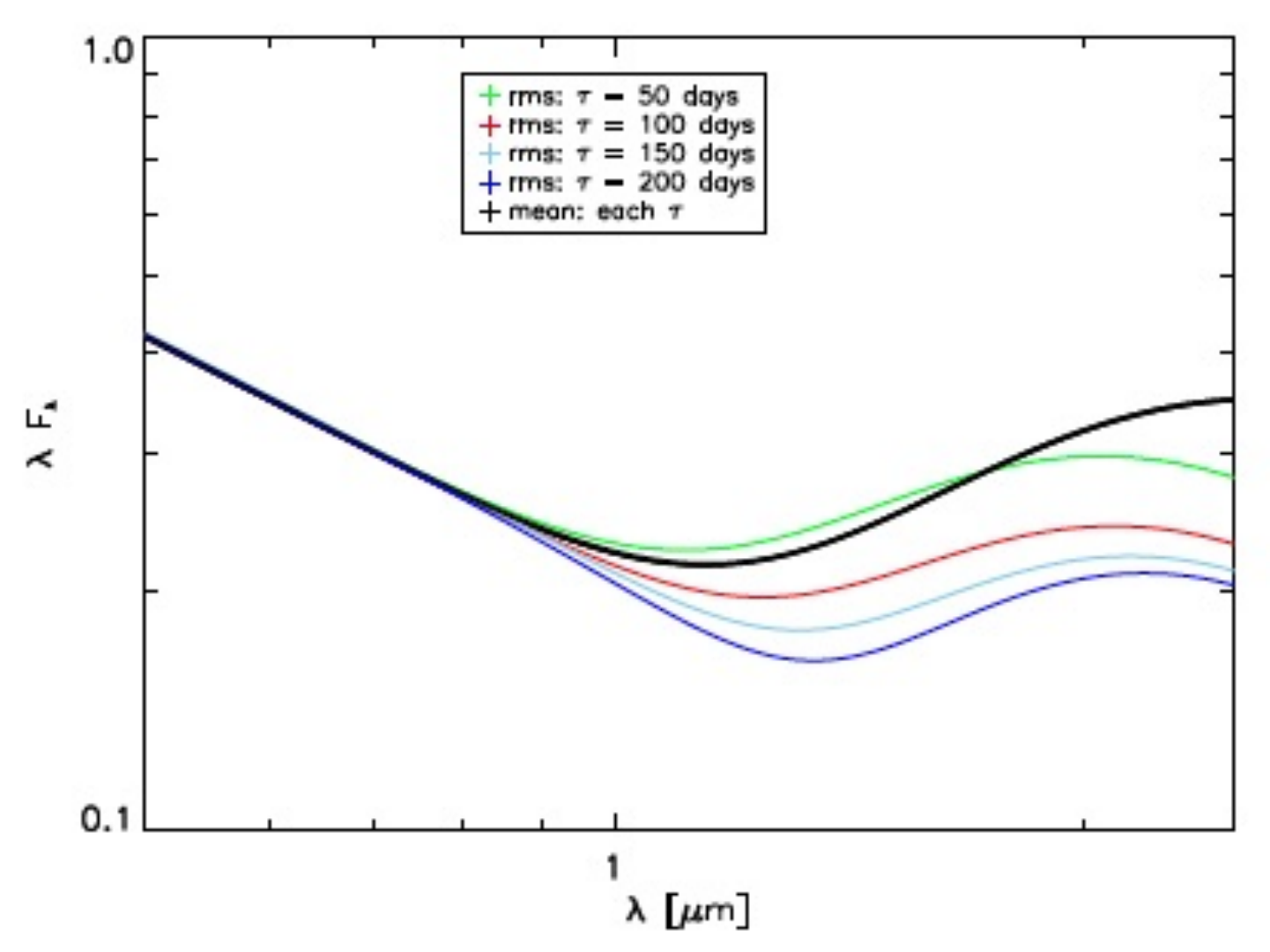}
  \includegraphics[scale=0.4]{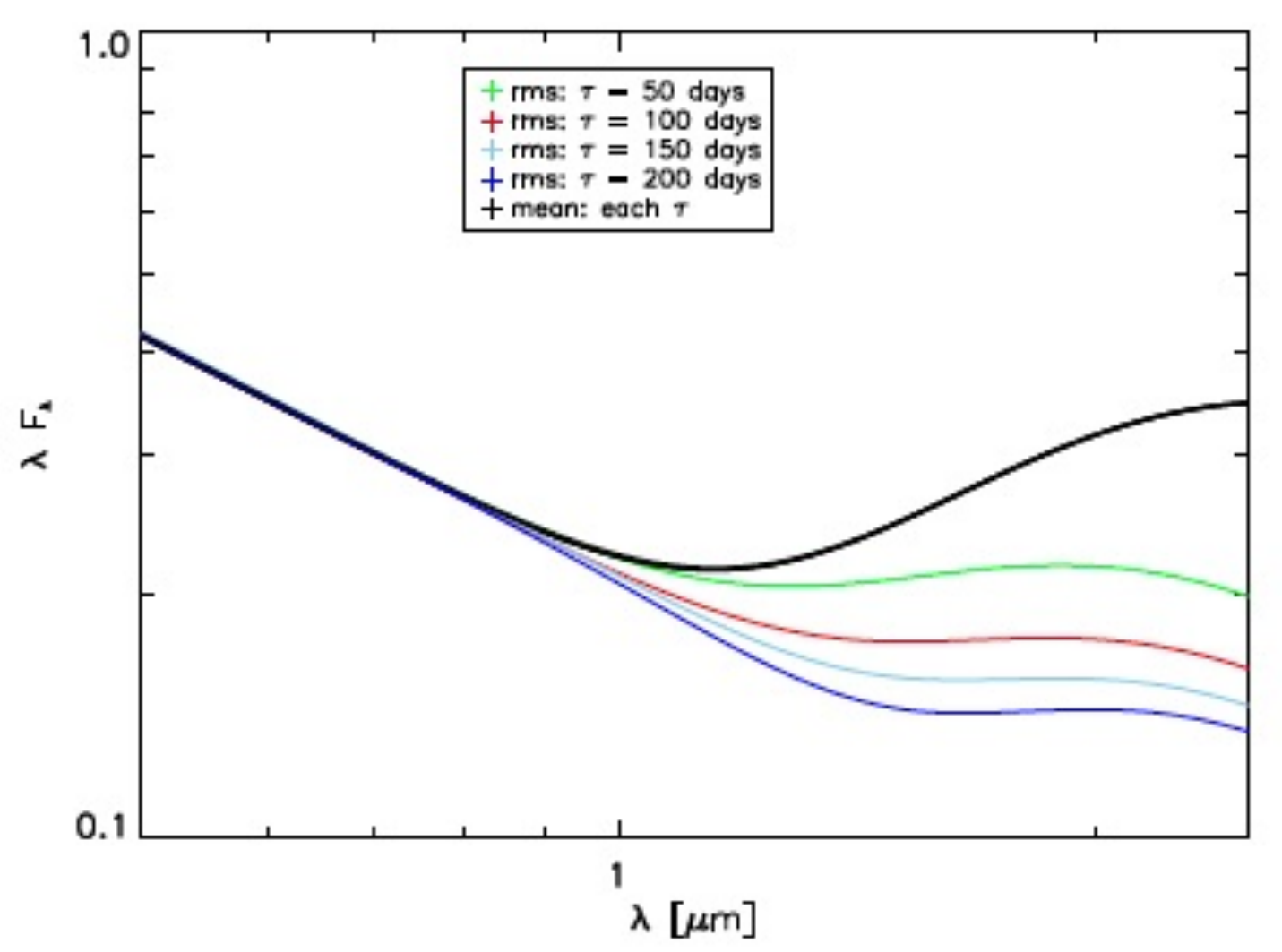}
}
\caption{\label{demcsimulation} Simulations of the mean (black line) and rms spectra (coloured lines) with the \demc~formalism assuming a secondary variable component at near-IR wavelengths, which is produced by the accretion disk emission. The rms spectra are simulated for four different dust response times ($\tau = 50, 100, 150$ and 200~light-days, indicated by the green, red, light blue and dark blue solid lines, respectively) and three different values for the dust processing efficiency ($\nu = 1, 0.8$ and 0.5, shown in the left, middle and right panel, respectively).}
\end{figure*}

For the simulations, we assumed a fixed campaign length of 1000~days and modelled the input light-curve with stochastic short-term variability overlaid with a roughly sinusoidal long-term variability with a period length of $\sim 400$~days. As the light-curves presented in \citet{Miller22} show, this is a reasonable assumption for the variability pattern of the accretion disk in Mrk~876, which exhibits a quasi-oscillatory trend over a time period of about three years. Their data cover two peaks and two troughs of the long-term variability pattern, indicating roughly a period length of about twice the dust response time. Then, we chose response times of the hot dust emission relative to the variable (optical) accretion disk flux of $\tau = 50, 100, 150$ and 200~light-days. In addition, we varied the processing efficiency of the dust and assumed values of $\nu = 0.5, 0.8$ and 1, with the latter producing the maximum possible dust flux variability. Fig.~\ref{demcsimulation} shows our results. It is clear that this model of two variable near-IR components can qualitatively reproduce the observations. The strongest depression of the hot dust variations is achieved for a combination of response time that gives maximal weakening of the total near-IR variability signal, i.e. half the period length and so $\sim 200$~light-days, and an intrinsically reduced fractional variability of the hot dust emission due to a low processing efficiency. As the accretion disk flux contribution to the total near-IR flux decreases with wavelength (see Fig.~\ref{irsed}), so does its weakening effect on the dust variability signal and the dust blackbody can always dominate the rms spectrum at the longest wavelengths. Then, if a secondary variable near-IR component is indeed present at $\sim 2~\mu$m and it is the accretion disk, the implied outer radius is $r_{\rm out} \sim 4300\,r_{\rm g} \sim 56$~light-days. Such an accretion disk is likely to harbour dust, which can contribute to the total blackbody emission. We will argue this case further in Section \ref{seconddust}. Finally, we note that a reduced value of $\nu$ in our simulations has a similar effect as adding a constant dust component, i.e. it decreases the dust fractional variability. Therefore, it is conceivable that the observed rms spectrum of Mrk~876 is a combination of both scenarios described above.

\section{The dust structure in Mrk~876}

It is of high interest to understand the chemical composition and grain size distribution of the circumnuclear dust in AGN, which can ultimately reveal how cosmic dust forms and evolves in different environments. If we can also constrain where the dust forms, we can assess its relationship to and influence on the other central structures of the AGN, such as, e.g., the BLR and accretion disk. As \citet{L19} showed, a spectroscopic near-IR reverberation mapping campaign that uses cross-dispersed data covering a very wide wavelength range is ideally suited to investigate these questions. In the following, we discuss our results for Mrk~876 with a focus on how the combination of a luminosity-based dust radius with a contemporaneous dust response time can constrain the astrochemistry of the dust (Section \ref{chemistry}), how the simultaneous measurement of the dust temperature and dust radius can shed light on the dust geometrical structure (Section \ref{dustywall}), and the potential of a near-IR variable (rms) spectrum to unravel the structure of accretion disks in AGN (Section \ref{seconddust}). Finally, we relate our results to AGN studies at other wavelengths and develop a new paradigm that connects AGN and protoplanetary discs (Section \ref{passivedisc}).

\subsection{The astrochemistry of the hot dust} \label{chemistry}

From the radiative equilibrium relationship (Eq. \ref{Stefan-Boltz}) it is clear that if estimates of the dust temperature $T$ and irradiating luminosity $L_{\rm uv}$ are available and the dust radius can be measured by an independent method, the dust emissivity parameter $Q^{\rm em}$, which strongly depends on the dust chemical species and grain size, can be constrained. Independent measures of the dust radius can come from, e.g., the dust response time measured through reverberation mapping, a geometric measurement based on near-IR interferometry or a polarimetric estimate of the location of the scattering region. 

Our near-IR spectroscopic data does not extend over a time period long enough to be able to derive a reliable dust response time for Mrk~876. However, the {\it WISE} monitoring data for the years \mbox{2010-2018} analysed by \citet{Lyu19} overlap with our campaign and the $W1$~band (3.4~$\mu$m) is expected to sample the same hot dust component by covering its flux close to the SED maximum (see Fig.~\ref{irsed}). \citet{Lyu19} estimated a dust response time of $\tau = 297\pm17$~light-days (observer frame) for the $W1$ band, assuming no corrections (see their Fig.~5). Adjusting the initial guesses for their fitting routine based on cross-correlation function results, the revised value was $\tau = 372\pm63$~light-days (see their Table~4). Their result is consistent with the distance to the equatorial scattering region of $254\pm39$~light-days (rest-frame) estimated by \citet{Afa19} based on spectropolarimetric observations from 2015. Furthermore, the dust response time in Mrk~876 appears to remain relatively constant. From an optical/near-IR photometry campaign during the years \mbox{2003-2007}, \citet{Min19} obtained values similar to \citet{Lyu19} of $\tau = 320\pm34$, $327^{+42}_{-36}$ and $327^{+41}_{-35}$~light-days (observer frame), assuming three different power-law spectral slopes for the accretion disk spectrum when decomposing the flux in the $K$~band. 

This independent measure of the dust radius of \mbox{$\sim 300$}~light-days is smaller than all average dust radii we estimated from thermal equilibrium considerations. The least discrepancy is found relative to the case of a pure blackbody (\mbox{$\langle R_{\rm d,lum} \rangle = 469\pm30$}~light-days), which corresponds to dust composed of large grains. The average luminosity-based radius for dust of small grains is much larger and exceeds the dust response time by a factor $\ga 10$ and $\ga 16$ for carbon and silicate dust, respectively. Therefore, it seems that the hot dust in Mrk~876 is dominated by grains with sizes of at least a few $\mu$m, as found for NGC~5548 by \citet{L19}. However, contrary to the case of NGC~5548, the luminosity irradiating the dust in Mrk~876 appears to be only a fraction of the bolometric luminosity that we see. Alternatively, if the dust is exposed to the total bolometric luminosity, something must reduce its temperature. In this case, it is worth noting that the response-weighted dust radius predicts a dust temperature of $T \sim 1700$~K, which is close to the sublimation temperature for carbonaceous dust \citep[$T_{\rm sub} \sim 1800 - 2000$~K;][]{Sal77}. Carbonaceous dust was found to be the dominant species in the hot dust of NGC~5548.

It was previously postulated that the circumnuclear dust in AGN is dominated by large grains \citep[e.g.][]{Laor93, Mai01b}. The spectroscopic near-IR monitoring campaigns on NGC~5548 and Mrk~876 have shown that this is indeed the case. Furthermore, they put an important lower limit on the dust grain size of a few $\mu$m, which suggests that the hot dust is assembled in a disk-like structure. When the dust grains are very small, their coupling to the ambient gas is strong and Brownian motion dominates the collision rate. Brownian-motion driven agglomeration can efficiently grow $\mu$m-sized dust particles, but the induced collision velocity decreases with increasing aggregate mass. Then, as dust grains grow they increasingly decouple from the gas and differential settling becomes the main driver for grain collisions and so further grain growth \citep{Weid77, Blum08}. Therefore, large dust grains are expected to be found mainly in the midplane of a disk-like structure. If the gas is turbulent, we expect it to drive diffusion of dust in the vertical direction, which will oppose the effect of vertical settling and give the disk-like dust structure a scale height.

\subsection{The enlarged dust-free inner hole} \label{dustywall}

The lower dust reponse time in Mrk~876 relative to the dust radius estimated from radiative equilibrium considerations contrasts with the findings of \citet{L19} for NGC~5548, where the two dust radii were similar. However, it is in line with general comparisons between interferometric dust radii and those determined by photometric reverberation campaigns; the former are found to be larger than the latter by a factor of $\sim 2$ \citep{Kish09, Kish11, Kosh14}. Since interferometry effectively measures the luminosity-weighted radius, this result can be understood if one assumes that response-weighted radii are dominated by dust located at the smallest radii, whereas luminosity-weighted radii are dominated by dust grains with the highest emissivity, i.e. the hottest and largest grains. Due to their lower heat capacity, the highest temperature large grains can attain will always be lower than the highest temperature small grains can be heated to, and so their emission will come from further out. We can easily rule out this scenario for Mrk~876 since the blackbody temperatures in the mean and rms near-IR spectrum are similar.

Alternatively, \citet{Kaw10, Kaw11} ascribed the observed difference between response- and luminosity-weighted dust radii to a concave (or bowl-shaped) dust geometry. Due to the anisotropy of the accretion disk emission, dust located at larger viewing angles relative to the irradiating UV/optical flux will see it reduced by a factor $\cos \theta$, with $\theta$ the angle between the accretion disk rotation axis and the location of the dust. The interferometric radii would then be dominated by the region containing the bulk of the dust mass, which in this geometry is located further out, whereas reverberation would simply sample the smallest and so most variable radii. In this scenario, the concave shape of the dust structure is caused by the assumption of a constant maximum dust temperature, taken to be the dust sublimation temperature, which is found further away the higher the accretion disk luminosity appears to the dust. Our observations for Mrk~876 can be well explained with this scenario. The variable and mean near-IR spectra yield similar dust temperatures, albeit below the sublimation temperature. Then, in order to make the luminosity-based dust radius consistent with the dust response time, we require that the variable dust sees the bolometric luminosity reduced by a factor of $\sim 4$. If our viewing angle to the accretion disk is $\theta = 0^\circ$, as assumed in our calculations of the accretion disk spectrum (Section \ref{lumradius}), a variable dust component in a structure inclined at $\theta \sim 75^\circ$ and so closely aligned with the plane of the accretion disk would fulfill this requirement. The large dust deficit observed in the rms spectrum relative to the mean spectrum (Section \ref{rms}) would then be due to the fact that the total emission is dominated by the much larger dust mass located further out, which appears constant since it varies on much longer timescales than our campaign is sensitive to. 

An indication that the accretion disk illumination anisotropy considered by \citet{Kaw10, Kaw11} affects low- and high-luminosity AGN differently was also found by \citet{Min19}. Their sample, which included NGC~5548 and Mrk~876 at the low- and high-luminosity ends, respectively, and spanned about four orders of magnitude in luminosity, showed a best-fit slope for the logarithmic reverberation dust radius versus optical luminosity relationship of $\sim 0.4$, i.e. shallower than the slope of 0.5 predicted by thermal equilibrium considerations. Interestingly, using the dust response time and average accretion disk luminosity for NGC~5548 determined by \citet{L19} and the values of these parameters for Mrk~876 from our study we derive also a slope of $\sim 0.4$. Clearly, it would be worthwhile to further populate the undersampled top end of the relationship between dust radius and optical continuum luminosity in order to investigate how the dust irradiation pattern depends on accretion disk luminosity.

But one of our most important results cannot be explained solely with the scenario of \citet{Kaw10, Kaw11}, although the implied flared disk-like structure for the hot dust helps. Similarly to what \citet{L19} observed in NGC~5548, the hot dust in Mrk~876 is not close to its sublimation temperature. This finding implies the presence of an enlarged dust-free inner region in AGN and so a luminosity-invariant inner edge of the torus, i.e. a ``torus wall''. Since the sublimation temperature for carbon dust is higher than for silicate dust ($\sim 1900$~K and $\sim 1400$~K, respectively), the extent of this `inner hole' would be considerably larger if indeed the former chemical species dominates the dust composition. Such a dusty wall was confirmed for NGC~4151 with several epochs of both interferometric and reverberation dust radius measurements \citep{Kosh09, Pott10, Schnuelle13}. This paradigm for the AGN dust structure is similar to the standard picture for protoplanetary disks around young stars. Interferometry finds in many of these disks that the inner dust radius is significantly larger than the dust sublimation radius. This cavity, which is commonly referred to as the 'inner hole', is believed to be filled with gaseous disk material that can change its optical thickness and thus influence the location of the puffed-up 'wall' or 'inner rim' of the dusty, flared and passively illuminated protoplanetary disk \citep{Monnier05, Dull10}. Our result that circumnuclear AGN dust is composed of relatively large grains was found also for protoplanetary disks \citep{vanBoekel04, Kospal20}. We will explore in more detail the similarity between these two classes of astrophysical objects in Section \ref{passivedisc}.

\subsection{A secondary hot dust component} \label{seconddust}

\begin{figure}
\centerline{
\includegraphics[scale=0.3]{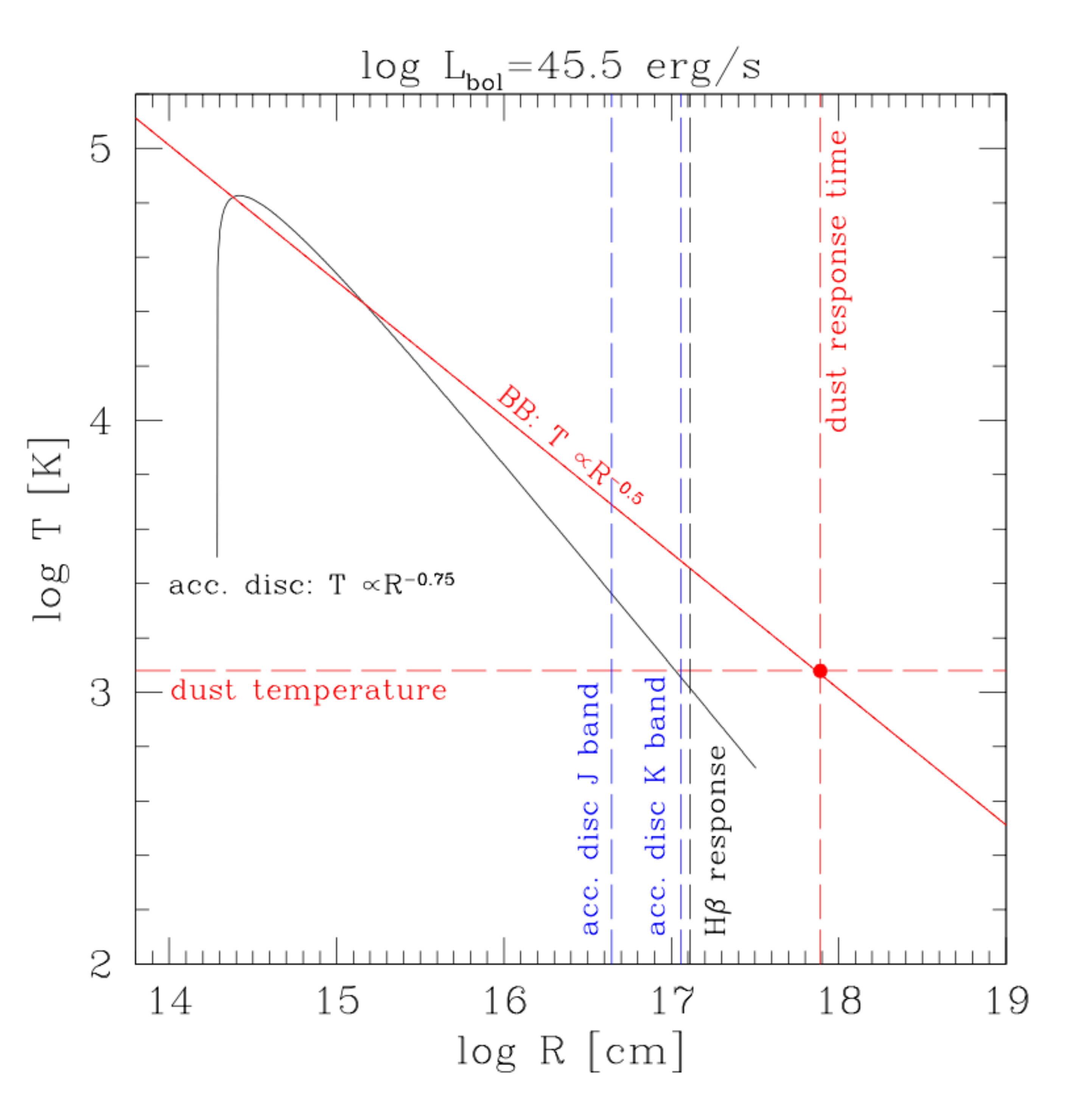}
}
\caption{\label{tempradius} The temperature-radius relationships for the hot dust (red curve) and the accretion disk illuminating it (black curve). They correspond to a bolometric luminosity of \mbox{$\log L_{\rm bol} = 45.5$~erg~s$^{-1}$}, which we derived from ``calibrating'' the relationship for the hot dust to its observed temperature in the rms near-IR spectrum and the rest-frame dust response time ($T = 1200$~K and $R_{\rm d,rev} = 300$~light-days, shown as the red horizontal and vertical dashed lines, respectively). The blue dashed lines indicate the accretion disk reverberation radii of $\sim 20$ and $\sim 45$~light-days estimated by \citet{Miller22} for the $J$~and $K$~bands, respectively. The black vertical dashed line indicates the H$\beta$ BLR response time of $\sim 50$~light-days measured by \citet{Bao22}.}
\end{figure}

In Section \ref{rms}, we showed that the observation of a reduced dust emission in the rms near-IR spectrum of Mrk~876 relative to its mean spectrum can also be explained by the accretion disk spectrum extending well into the near-IR. The implied large disk radii were $\sim 25$ and $\sim 56$~light-days in the $J$ and $K$~bands, respectively. These values are similar to estimates from a contemporaneous accretion disk reverberation mapping campaign. \citet{Miller22} recently presented the accretion disk analysis using optical photometric monitoring in several bands ($u$, $g$, $r$, $i$, and $z$) conducted between 2016 Mar and 2019 May, including our data set presented in Section~\ref{photometry}. They found that the response times depend on the wavelength as expected for a standard thin disk. Then, by extrapolating their optical response times to the near-IR, they estimated outer disk radii of $\sim 20$ and $\sim 45$~light-days in the $J$ and $K$~bands, respectively. Given these extents, it is of interest to understand if they are large enough to potentially harbour dust.

We can estimate the location of the accretion disk dust, if it exists, as follows. First, we used the simultaneous measurement of the dust temperature in the rms spectrum and the dust response time for the variable hot dust component together with the emissivity law we determined in Section \ref{chemistry} to ``calibrate'' the temperature-radius relationship for the dust and thus to estimate the bolometric luminosity that produces it. We show this relationship, which is that of a blackbody and so has the form \mbox{$T \propto R^{-0.5}$}, in Fig.~\ref{tempradius} (red solid line). As already discussed in Section \ref{dustywall}, the corresponding total bolometric luminosity that seems to irradiate the dust is \mbox{$\log L_{\rm bol} \sim 45.5$~erg~s$^{-1}$}, which is a factor of $\sim 4$ lower than the average bolometric luminosity of $\log L_{\rm uv} = 46.09 \pm 0.03$~erg~s$^{-1}$ estimated from fits to the near-IR spectra. Assuming that this irradiating luminosity is produced by an accretion disk, we then calculated the temperature-radius relationship for it, which is of the standard form $T \propto R^{-0.75}$ and so steeper than that for the hot dust (black solid line). These curves immediately show that dust is expected to condense out in the accretion disk gas at much smaller radii than those possible for gas exposed to the full bolometric luminosity. Then, if this dust was driven off the accretion disk, e.g., by a dusty outflow, it would be quickly destroyed at considerable heights \citep{Czerny17}. For the observed dust temperature, we expect the accretion disk of Mrk~876 to become dusty at radii $\ga 40$~light-days. This secondary hot dust will not necessarily be externally illuminated and so reverberate in the classical sense. Instead, if it is composed of $\mu$m-sized grains, it will be well-coupled to the gas and vary on the same timescales. Dust with a variability on such short time-scales is likely to appear constant in our rms near-IR spectrum. 

An additional argument for the reality of a secondary hot dust component in Mrk~876 can also be derived from the total observed dust luminosity. The values listed in Table~\ref{lumradiustab} (column (4)) give an average of \mbox{$\log L_{\rm d} = 45.55 \pm 0.02$~erg~s$^{-1}$}, which, if interpreted as being produced by a single dust component (e.g. with the geometry proposed by \citet{Kaw10}), implies a dust covering factor of unity given the ``calibrated'' bolometric luminosity irradiating it. Given that the hot dust is optically thick to the UV/optical accretion disk luminosity, such an AGN would be an obscured, Compton-thick one. On the other hand, if we assume an additional dust component in the mean spectrum, the rms spectrum tells us that the variable dust component makes up only roughly a third of the total observed dust flux and so has a covering factor of $\sim 30\%$. We note that by coincidence we would estimate a similar value for the covering factor also from the mean spectrum alone, since both the bolometric luminosity illuminating the torus dust and the dust luminosity would be overestimated by a similar factor. Then, if the dust deficit in the rms spectrum appears exacerbated due to the influence of the variable accretion disk (and possibly its associated dust), as discussed in Section~\ref{rms}, the torus dust covering factor would be higher.

The first tentative evidence for dust in the accretion disk was found for NGC~5548 by \citet{L19}. If it can be shown that such dust is ubiquituous in AGN, it would provide a possible connection to the BLR within models for radiatively accelerated dusty outflows launched from the outer accretion disk \citep[e.g.][]{Elitzur06, Czerny16, Czerny17, Esser19}. In this case, carbon dust would provide the largest opacity and lead to an inflated disk structure \citep{Baskin18}. Support for this scenario in Mrk~876 comes from the remarkable agreement between the location we estimated for the accretion disk dust and the H$\beta$ BLR radius of $\sim 50$~light-days measured by \citet{Bao22} for their reverberation mapping campaign conducted between 2016 Dec and 2018 May. Disk-like dust structures in AGN have started to be revealed also by interferometric observations achieving an unprecedented high spatial resolution in the mid-IR \citep{Isbell22}.

\subsection{Are AGN and protoplanetary disks similar?} \label{passivedisc}

Large accretion disks that extend well beyond the self-gravity radius are now routinely inferred for AGN by accretion disk reverberation studies \citep[see review by][]{Cack21}. Evidence for the presence of dust in these disks would imply even greater radii, thus exacerbating the stability issue. A possible solution and an alternative explanation for our finding of a significantly reduced bolometric luminosity of the accretion disk in Mrk~876 as seen by the dust is that it is a passive disk. A passively illuminated disk, such as a protoplanetary disk, has a temperature-radius relationship indistinguishable from that of a standard accretion disk that produces its own radiation via gravitational energy release, i.e. it is of the same form $T \propto R^{-0.75}$ \citep[e.g.][]{Cack07}. Accretion disk flux variability in AGN is readily interpreted as reprocessed emission from a central, highly variable, hard X-ray source since the observed time-scales are much shorter than the viscuous time-scale, i.e. the time it takes for a change in accretion rate to trigger a change in disk temperature and so flux. The passively irradiated flux is usually assumed to be only a small fraction of $\sim 10-20\%$ of the total disk emission \citep{Starkey16, Gardner17}. The value of $\sim 25\%$ that we estimated for Mrk~876 is close to this range. An advantage of the passive disk scenario over the $\cos \theta$-effect in a bowl-like geometry, such as the one proposed by \citet{Kaw10}, is that it predicts the onset of dust in the accretion disk around the H$\beta$ BLR radius; if only an anisotropic accretion disk emission as seen by the dust is assumed, the accretion disk itself would have the total (unreduced) bolometric luminosity and so be hotter, leading to an onset of dust much further out. We also note that the relative scaling we obtained between the mean and rms near-IR spectra in the optical regime of a factor of $\sim 6$ (Section \ref{rms}) is similar to the factor of $\sim 4$ reduction in total bolometric luminosity required by the variable dust. Therefore, it could well be that the dust is irradiated only by the variable, reprocessed disk flux.

Passive disks can be distinguished from standard accretion disks, since, as a consequence of vertical hydrostatic equilibrium, they are flared, with the disk relatively thicker at larger radii. The vertically isothermal flared disk diverges from the flat disk solution at intermediate radii and approaches $T \propto R^{-3/7}$ at large radii, i.e. it is very similar to a single blackbody \citep[e.g.][]{Kenyon87}. The dust is crucial for the disk thermodynamics (and for planet formation), but represents only $\sim 1\%$ of the total mass and is easier to observe than the gas. Self-gravity is generally only significant in very massive disks and protoplanetary disks are often not massive enough to become gravitationally unstable. It is then conceivable that in Mrk~876 the sum of both temperature-radius profiles displayed in Fig. \ref{tempradius} (black and red solid lines) represents a single entity, namely, a passively illuminated, flared and dusty disk. A wind off this disk could then be the origin of the BLR.

\section{Summary and conclusions}

We have conducted the first spectroscopic near-IR monitoring campaign on Mrk~876, one of the intrinsically most luminous AGN in the nearby Universe. Our cross-dispersed spectroscopy can measure dust temperatures with high precision, which allows us to derive the luminosity-based dust radius for different grain properties. When comparing it to an independent measure of the dust radius, we can then constrain the astrochemistry of the hot dust. Furthermore, we can construct the variable (rms) near-IR spectrum over a relatively large wavelength range. 

Our main results can be summarised as follows.

\vspace*{0.2cm}

(i) Assuming thermal equilibrium for optically thin dust, we find that the luminosity-based dust radii are larger than the dust response time obtained by a contemporaneous photometric reverberation mapping campaign, with the least discrepancy (of a factor of $\sim 2$) found relative to the result for a wavelength-independent dust emissivity law, i.e. a blackbody, which is appropriate for grains of relatively large sizes (of a few $\mu$m). This result can be well explained by a flared, disk-like structure for the hot dust, whereby the anisotropy of the accretion disk emission causes a decrease in illumination (by a factor of $\sim 4$ in our case), as first proposed by \citet{Kaw10, Kaw11}.  

(ii) The near-IR variable (rms) spectrum tracks the accretion disk spectrum out to longer wavelengths (of $\sim 1.2~\mu$m) than the mean spectrum (of $\sim 1~\mu$m) due to a reduced dust emission in the former. The implied outer accretion disk radius is much larger than the self-gravity radius and consistent with the extrapolated results from the contemporaneous, multi-band optical accretion disk reverberation mapping campaign of \citet{Miller22}. The flux variability of the hot dust with respect to the accretion disk is reduced by a factor of $\sim 3$ and could be due to either the presence of a secondary hot dust component in the mean spectrum or the destructive superposition of the dust and accretion disk variability signals or some combination of both. 

(iii) The large extent of the accretion disk, which is likely to harbour dust in its outer regions, and the low bolometric luminosity as seen by the hot dust can also be explained if we assume that AGN disks are similar to protoplanetary disks around young stars. A passively illuminated, flared and dusty disk would naturally provide a single, continuous structure for the temperature-radius profile determined by AGN accretion disk reverberation studies ($T \propto R^{-3/4}$) at small radii and that of the hot dust blackbody ($T \propto R^{-1/2}$) determined here at large radii.

\bibliography{references}{}
\bibliographystyle{aasjournal}

\begin{acknowledgments}
HL is indebted to Brad Peterson for suggesting the science target and leading the infrared observing proposals and to Peter Abraham, Agnes Kospal and Hubert Klahr for generously sharing their knowledge about protoplanetary disks. HL and JAJM thank Michael Fausnaugh for help with applying his \mapspec~routine to the near-IR spectroscopy. HL acknowledges a Daphne Jackson Fellowship sponsored by the Science and Technology Facilities Council (STFC), UK. J.A.J.M acknowledges the support of STFC studentship ST/S50536/1. HL, JAJM and MJW acknowledge support from STFC grants ST/P000541/1 and ST/T000244/1. KH and JVHS  acknowledge support from STFC grant ST/R000824/1. ERC acknowledges the support of the South African National Research Foundation.
\end{acknowledgments}

\vspace{5mm}
\facilities{Gemini(GNIRS), LCOGT(optical)}

\end{document}